

\def\lrpartial{\mathrel{\partial\kern-.75em\raise1.75ex\hbox
{$\leftrightarrow$}}}

\line{
\hfill ULB-TH 03/94}
\line{\hfill gr-qc/9404026}
\line{
\hfill April 1994}
\bigskip\bigskip
\centerline  {\bf SOURCE VACUUM FLUCTUATIONS OF BLACK HOLE RADIANCE}
\bigskip\bigskip
\centerline{F. ENGLERT$^{a,b}$\foot{E-mail: fenglert at
ulb.ac.be}, S. MASSAR$^a$\foot{Boursier IISN. E-mail: smassar at
ulb.ac.be} and R. PARENTANI$^c$\foot{E-mail: renaud at
vms.huji.ac.il}} \medskip
 \centerline{\item {$a)$}{\it Service de Physique Th\'eorique}}
\centerline{\it Universit\'e Libre de Bruxelles, Campus Plaine, C.P.225 }
\centerline{\it Boulevard du Triomphe, B-1050 Bruxelles, Belgium}
\medskip
\centerline{\item{$b)$}{\it School of Physics and Astronomy}}
\centerline{\it Raymond and Beverly Sackler Faculty of Exact Sciences}
\centerline{\it Tel-Aviv University, Ramat-Aviv, 69978 Tel-Aviv, Israel}
\medskip
\centerline{\item{$c)$}\it Department of Theoretical Physics}
\centerline{\it The Racah Institute of Physics}
\centerline{\it The Hebrew University of Jerusalem, Givat Ram Campus}
\centerline{\it Jerusalem 91904, Israel}
\bigskip
\bigskip
\noindent
{\bf Abstract}

The emergence of Hawking radiation from   vacuum
fluctuations
 is analyzed
  in  conventional field theories and their  energy content is defined
through the
 Aharonov weak value concept. These fluctuations travel in flat
space-time
 and carry
  transplanckian energies sharply localized on cisplanckian distances. We
argue that
 these features
  cannot accommodate
   gravitational nonlinearities. We suggest that
the very emission of
 Hawking photons from
  tamed vacuum fluctuations requires the existence of an
exploding set of  massive fields. These considerations corroborate some
conjectures of Susskind
 and may prove relevant
  for the back-reaction problem and for the unitarity
issue.
 \vfil \eject

\noindent
{\bf 1. Introduction}

The remarkable discovery
 by Hawking$^{[1]}$  of the thermal radiation of an incipient black-hole
has raised many questions
 about quantum gravity but has  as yet not delivered   conclusive
answers. The very existence
 of the radiation confirmed
  the Bekenstein conjecture relating the
area of the event horizon to
 entropy$^{[2]}$ but did not yield the identification
of the quantized  matter-gravity
 states building up this entropy. The back-reaction of the
radiation on the metric  should lead to the evaporation of the
hole but its precise mechanism is
 far from being understood and the end point of the evaporation
itself poses in an acute way the
 consistency of quantum physics with general relativity. Is
unitarity violated within our
 universe as initially suggested by Hawking$^{[3]}$ or does the
planckian black hole turn into
 a (infinitely?) long lived remnant$^{[4]}$ correlated to the
distant radiation? This question
 is related to the value of a possible additive constant to
the area entropy which would count the remnant degeneracy; attempts to
understand the nature of the
 constant$^{[5]}$ have been made but no definite conclusion has
been reached. Other more
 revolutionary attempts to save unitarity through a  breakdown of
large scale physics have
 been proposed$^{[6],[7]}$ but remain conjectural.

The heart of the difficulty
 lies in the quantum back-reaction. The semi-classical treatmemt, whereby
the expectation value of the
 energy-momentum tensor of the radiation is taken as the source to the
classical Einstein equations$^{[8]}$,
 is questionnable in view of the importance of the
fluctuations. In particular the fact
 that the original derivation of the emission
process$^{[1]}$ requires vacuum
 fluctuations of frequencies much higher than the Planck scale may
change completely the nature of the back-reaction. The
occurrence of these
 transplanckian\foot{A picturesque adjective we heared from 't Hooft at a
meeting.} frequencies,
 although
   consistent in the  free field description of the vacuum, poses a
moot problem when
 gravitational
  nonlinearities are introduced$^{[9]}$. To clarify the issue, it
would be helpful
 to have a complete
  description of the pair creation process generating Hawking
quanta out of the free
 field vacuum. However despite the earlier work of Unruh$^{[10]}$ and
Wald$^{[11]}$, and the
 more recent clarifications
  of Parentani and  Brout$^{[12]}$, the explicit
history of the correlated pairs
 for  free massless fields have not been brought to
light. We shall display the
  history in a   simplified collapse situation which we think contains
the relevant
 physical properties.
 Our result is that nearly all the quanta are generated from
dipolar vacuum fluctuations
 in flat Minkowski space-time with transplanckian frequencies sharply
localized at cisplanckian
 distances. For s-waves
  these dipoles are spherical;
   they travel from past
light-like infinity towards the
 centre of the star in an essentially flat background and then
separate. Their energy content
 is displayed following reference [13] where a systematic analysis of
the energy momentum tensor in
terms of weak value$^{[14]}$ is given. The positive energy pole flows
towards  an eventual horizon
while the negative pole, composed of a positive energy core followed by
an oscillatory
 tail with overcompensating
  negative energy, crosses
  the star surface, loses its tail
and gets converted into a
 Hawking quantum by reducing
  its core energy to confront the gravitational
background outside the star.
 The history of higher angular momenta modes is essentially the same,
except that after crossing the
 star's surface most of the outgoing modes get reflected. The result is
a thermal distribution close to
 equilibrium in the vicinity of the horizon. This analysis will lead
us to the  conclusion that while
 independent arguments suggest quite convincingly that   Hawking
radiation does occur, the   mechanism
 that  produces it, taking gravitational effects into   account,
cannot be realized
 out of the conventional
  cisplanckian physics, except for the emission of the very
first quanta.

We shall suggest that the  emission process requires, in addition to the
massless fields generally considered, an exploding set
of massive fields such as
 those encoded in weakly interacting closed string theories. Such
structure, in four dimensional
 space-time, exhibit a high temperature phase transition when the
energy stored in the massless
 modes exceeds a critical energy density, the excess energy
condensing into large massive
 strings with huge entropy. This provides a dynamical mechanism to
reduce transplanckian frequencies
 to planckian and cisplanckian ones. Further investigations
along these lines may prove valuable
 for uncovering some features of quantum gravity and maybe
of the string theory approach itself. They could shed light on the
back-reaction process and on the unitarity issue.

The presentation is as follows. In
 section 2, the conventional derivation of the Hawking
radiation is reviewed for a simple idealized collapsing star
in the restricted space-time available
 to external observer. In this way, we avoid, in the
subsequent discussions, unnecessary
 references to an eventual horizon. This formalism is
used in section 3 to uncover the
 history of s-wave vacuum fluctuations giving birth to
Hawking quanta in absence of
 back-reaction. Their energy content is obtained.
The shortcomings of  this history,  generalized to include higher
angular momentum waves, are
 discussed in section 4. A possible remedy based on elements
contained in closed string
 theories is presented.
 \bigskip  \noindent {\bf 2.  Pair
creation in the external observers space-time.}

The Hawking radiation due
 to a collapsing star is often analysed in the framework of the
global space-time background
 generated by the collapse, extending through the future
event horizon up to the classical
 singularity. We shall find it convenient to  phrase it in the
restricted space-time available to
 the outside observer which is limited by the  horizon. Our
analysis will therefore not rely on
 the existence of the event horizon itself but rather on the
geodesic motion of the star at late
 Schwartzschild times. In this way no a priori assumption
about the formation of a horizon is
 needed and the analysis is easily extended to the case where
the star, after having followed for
 some time an ``asymptotic" geodesic motion, would   slow
down to rest or would bounce back$^{[6]}$.
  Hawking radiation would   still be emitted during
the geodesic motion$^{[12]}$ but its subsequent
 alterations would automatically contain the required
correlations to render the radiation process unitary.
 As will be later discussed, the
inclusion of gravitational effects  raises a fundamental difficulty
which   is essentially the same for a stopped collapse
 and for a real one.

Following the pioneering work of Unruh$^{[10]}$ and the
 more recent analysis of
reference $[12]$, we   consider a spherically symmetric
 star of mass $M$ idealized by a
shell of the same mass collapsing along the geodesic
 trajectories of the points located at the
surface of
 the star,
  that is   geodesic trajectories of a Schwartzschild geometry with mass
$M$. Tensions
 in the shell
  must be adjusted accordingly. This idealization simplifies the
mathematics without affecting qualitatively the conclusions.

For the external
 observer, in absence
  of back-reaction, space-time is limited to the shaded region of
the Penrose diagram of
 Fig.1 depicting the classical collapse of the star. This region can be
described by tortoise
coordinates outside the
 shell and by Minkowskian ones inside. Thus, outside,
$$ \eqalign {  &ds^2 =
\left(1- {2M\over r }\right) du\  dv - r^2
d\Omega^2 \cr
 &u= t-r^* \qquad v=t+r^*
  \cr &dr = (1- {2M\over r}) dr^* ,} \eqno (1)   $$
where $r$, understood as a
 function of $v-u$, is the ``radius" which measures the invariant surface
$4\pi r^2$ of a sphere. Inside the shell one may write
$$\eqalign{&ds^2= dU\ dV -r^2
  d\Omega^2 \cr &U=\tau - r \qquad V=\tau + r,} \eqno(2) $$
where  the general spherically
 symmetric solution of Einstein's equations  imposes that  $\tau$
be a function of $t$ only.

One can choose a single $(u,v)$
coordinate system covering the
 whole space-time available to the external observer which
coincides with the metric defined
 by Eq.(1) outside the shell. Keeping the ``conformal gauge"
of Eq.(1) in the two
dimensional $r,t$ subspace, this is entirely fixed by continuity of the
metric across the shell
 and by the continuity of $r$. Asymptotically close to the
Schwartzschild radius $2M$ one has
 $$  (1- {2M\over r}) \simeq  \exp\left({v-u \over 4M}\right)
\eqno (3) $$
 and at the surface of the star $v$ tends to a constant $v=v_\infty$.

In this asymptotic region,
 the trajectory of the shell is described, up to exponentially small
corrections,   by
$$v_s = v_\infty - 4M f
 \exp \left({v_\infty -u_s \over 4M}\right)    \eqno (4) $$
where f is a positive
 constant whose precise value depends on the initial conditions but
which remains of order
 one if the shell collapses from rest at distances large compared to
the Schwartzschild radius. In fact  a straightforward
computation of the geodesic
 motion yields   $f= 1/4$ for a   shell  at rest at $r=+\infty$.
The value of $f$
 then decreases  if the shell
  has  an initial velocity at $\infty$ and reaches zero in the
limit of a light-like shell.
The continuity of $ds^2$   yields, in the vicinity of the shell
surface, $$ dU= \lambda \exp
\left({v_\infty -u \over 4M}\right)  du \qquad dV = \lambda^{-1} dv
\eqno (5) $$
 where the constant $\lambda$
  is fixed by the continuity of $r$.  From Eqs.(1),(2) and
(3), we   write on the trajectory Eq.(4)
 $$2 dr = dV - dU =  \exp
  \left({v_\infty -u_s \over 4M}\right)  (dv - du).
\eqno(6) $$
 Using Eq.(5) and differentiating Eq.(4), Eq.(6) reduces to
$$\lambda^{-1} f =
\lambda - 1. \eqno(7)$$
 We see that
  $ \lambda$
  remains of
  order one in the above range of   initial conditions.

We have used here, inside the shell,  a time $\tau$ different from the
Schwartzschild
 time $t$. But
  one may synchronize the time inside and outside the shell by
parametrizing it everywhere by $t$; then inside, one writes
$$ d\tau = \sqrt{g^{in}_{00}(t)}
dt. \eqno(8)
$$ As $ dV + dU = 2 d\tau $, we now get from Eqs.(4) and (5), $$
\sqrt{g^{in}_{00}(t_s)}
= (2\lambda -1) \exp \left({v_\infty -u_s \over 4M}\right)=  (2\lambda
-1)g^{out}_{00}(t_s)  \eqno(9) $$
  where
$t_s$ is the
 synchronized
  time on the shell. Eq.(9) measures the redshift experienced by a
photon emitted
 from inside the
  star to infinity and crossing the shell at time $t_s$.  This
  redshift suffers a discontinuity across the shell
as a consequence of the geodesic motion imposed on the shell to mimic the
surface of a realistic collapsing star: if  the shell were at rest,  no
such discontinuity would appear.

We shall need the
 $u,v$ parametrization
  in the neighbourhood of the event horizon, not only
in the vicinity of the
 shell surface where it is close to the Schwartzschild radius, but also
deep inside the shell,
 where $r$ goes to zero. There, Eq.(5) still correctly defines $u$ in
terms of $U$ but not $v$
 in terms of $V$. The latter relation is indeed fixed by the surface of
the star at its intersection
 with the line $V=$constant and this point gets too far from the
Schwartzschild radius $2M$ to
 use the asymptotic form Eq.(3). Rather, $ (1- 2M/ r)$
gets closer to its limiting value
 at $r \to \infty$, namely one, and in this Minkowskian
limit  $V=v$. Thus, the second equation
 in Eq.(5) should be replaced by $dV=
\lambda^{-1}(v) dv$ where $\lambda^{-1}(v)$ is a slowly varying
function of $v$. This slow variation
 introduces  unessential complications due to the fact
that an incoming photon does not travel
 in an exactly flat space-time before entering inside
the shell.
 To avoid these,
  we shall therefore take $\lambda=1$ and independent of $v$. This
amounts to put
 $f=0$ in Eq.(7), and thus to consider the limit of a light-like shell.
Solving then Eq.(5)
 with $U=0$ at the
  event horizon ($u= \infty $) and $V=v=0$ at $r= U=0$, we
get
$$U= -4M \exp \left({v_\infty -u \over 4M}\right)
 \qquad V=  v \eqno(10) $$
and the trajectory $(u_0, v_0)$ of the centre of
the star, $2r=V-U = 0$, can be written as
$$ v_0 = -  4M \exp \left({v_\infty -u_0 \over 4M}\right)=
-A \exp \left({-u_0\over
4M}\right)  \eqno(11) $$ and, from Eqs.(2) and (10),
$$v_\infty = 4M. \eqno(12)$$
so that  $A =4M e   $.
 Eqs.(10),(11)
 and (12) and
 the value of $A$ would remain valid up to  factors of order (1) if
the light-like
 limit of the geodesic collapse were not taken. The reader may
verify that all
 relevant equations
  below would similarly only be affected by such factors. The
general classical collapse,
 as seen by an external observer, is depicted in Fig.2.

We now analyse the radiation
 emitted from the vacuum fluctuations of a massless scalar field
by the time-dependent metric.
 In sections 2 and 3, we consider only s-waves and neglect
the residual potential barrier.  The Heisenberg
scalar field operator
   rescaled by $r$ obeys then $ \partial_u \partial_v
\Phi=0$. It is   expanded
 into a complete set of   solutions $\phi_k$, that is
$$ \phi_k = f_k(u) + g_k(v)   \eqno (13) $$
such that
$$ (\phi_j\vert \phi_i) \equiv i\int_{\Sigma}
 [\phi_j^*\lrpartial_v \phi_i dv -
\phi_j^*\lrpartial_u \phi_i du ] = \delta_{ij}
 \; \hbox {or} \;\delta (i-j) \eqno(14) $$
where $\Sigma$ is an arbitrary
 Cauchy surface which does not cross the horizon (i.e. $ 0\leq r
<\infty , U <0)$, and the $\phi_k$
 vanish at $r=0$. From Eq.(11) such a complete set$^{[11]}$ (up to
hermitian conjugation) is
 $$\eqalignno{&\vert -\omega^{out})
 ={1\over \sqrt{4\pi \omega}}  \left[ \exp
(-i\omega u) -\Theta(-v) \exp (i4M\omega
 \ln {-v\over A})\right]&(15) \cr &\vert +\omega^{out})={1\over
\sqrt{4\pi \omega}}\Theta(v) \exp (-i4M\omega \ln {v\over A})&(16) \cr}$$
where  the frequencies $\omega$ span the
 positive real axis.  The out-modes $\vert
-\omega^{out})$ have positive frequencies
 with respect to the Killing vector on $\cal {I}^+$ so
that we write  $$\eqalignno{\Phi(u,v) =
 &\int_0^\infty  d\omega \left[\vert -\omega^{out})
a^{out}_{-\omega} + h.c.\right] &\cr +
&\int_0^\infty  d\omega \left[\vert +\omega^{out})
a^{out}_{+\omega} + h.c.\right].&(17)\cr} $$
 The creation operators $a^{out\
\dagger}_{-\omega}$ then creates quanta
 of energy $\omega $ on  $\cal {I}^+$ in a Hilbert space
$H_1$: these we call  Hawking ``photons".
 However the creation operators  $a^{out\
\dagger}_{+\omega}$ cannot
 be associated with well defined negative frequencies; they creates
states in a Hilbert space
$H_2$ orthogonal to $H_1$ which describes vacuum fluctuations
propagating towards the horizon.

A convenient complete set
of positive frequency in-modes on  $\cal {I}^-$ is
$$ \eqalign {
\vert \pm\omega^{in})={1\over
 \sqrt{8\pi \omega\ \sinh (\omega 4\pi M)}}[ &\Theta(v)    \exp
(\mp i4M\omega \ln {v\over A}
) \exp (\pm \omega 2\pi M) \cr + &\Theta(-v)\exp (\mp i4M\omega \ln
{-v\over A}) \exp (\mp \omega 2\pi M ),}\eqno(18)$$
 so that on $\cal {I}^-$, $\Phi(v)$ can be
written as
 $$ \Phi(v) = \int_0^\infty
 d\omega \left[\vert -\omega^{in}) a^{in}_{-\omega} +
\vert +\omega^{in}) a^{in}_{+\omega} + h.c.\right] .\eqno(19)$$

The Heisenberg vacuum $\ket 0$
 is annihilated by the operators $a^{in}_{1\omega}$. Identifying
Eq.(19) with Eq.(17) on $\cal
{I}^-$, we get the Bogoliubov transformation relating in- and
out- operators
 $$ \eqalign {  a^{in}_{+\omega}
  &=  \alpha_\omega a^{out}_{+\omega} - \beta_\omega
a^{out\ \dagger}_{-\omega}
\cr a^{in}_{-\omega} &=   \alpha_\omega a^{out}_{-\omega} - \beta_\omega
a^{out\ \dagger}_{+\omega}}\eqno(20) $$
where
$$ \alpha_\omega ={\exp
 (\omega 2\pi M)\over \sqrt{ 2\sinh (\omega 4\pi
M)}} \quad \beta_\omega
={\exp (- \omega 2\pi M)\over \sqrt{ 2\sinh (\omega 4\pi
M)}}.\eqno(21)$$
 Equivalently, if $\ket \Omega$ is the vacuum annihilated by
the operators
 $a^{out}_{1\omega}$,
  and  $\ket \Omega$ is the tensor product $ \ket {\Omega_1}
\ket {\Omega_2}$ of
 vaccua for $H_1$ and$H_2$,  $$\eqalignno {&U a^{out}_{1\omega}U^{-1}=
a^{in}_{1\omega}&(22)
\cr &\ket 0 = U \ket \Omega =U \ket {\Omega_1} \ket{ \Omega_2}
&(23)}$$ with
$$\eqalign {&U = \exp
   \int^\infty_0 d\omega \gamma_\omega [a^{out\  \dagger}_{-\omega}
a^{out\ \dagger}_{+\omega}
- a^{out}_{+\omega} a^{out}_{-\omega}]  \cr &\tanh \gamma_\omega =
\exp (- \omega 4\pi M) = {\beta_\omega \over
\alpha_\omega} .}\eqno(24)$$
Normal ordering the $U$ operator in Eq.(24) yields
$$\eqalignno{ \ket 0&=
 \langle\Omega\ket 0  \exp   \int^\infty_0 d\omega {\beta_\omega \over
\alpha_\omega} a^{out\
  \dagger}_{-\omega} a^{out\ \dagger}_{+\omega}\ket \Omega &(25) \cr
\langle
\Omega\ket 0 &= \exp -
\int^\infty_0 d\omega \ln \alpha_\omega \; \delta(0) \cr  &=\exp
-{T\over 2\pi}\int^\infty_0 d\omega \ln \alpha_\omega &(26)}$$
where $\delta(0)=T/2\pi$
 is the (infinite) time during which the black hole   radiates
quanta in a fixed background.

One sees from Eq.(25) that
 the state   describing a Hawking photon of frequency
$\omega$ $a^{out\ \dagger}_{-\omega}
 \ket {\Omega_1}$ has an   Einstein-Rosen-Podolsky  (EPR)
correlation with the state $a^{out\
\dagger}_{+\omega}  \ket {\Omega_2}$.
 Upon tracing  the pure state density matrix
$\ket 0 \bra 0$
over $H_2$, one recovers the outgoing s-wave thermal flux  at the Hawking
temperature
 $$T={1\over 8\pi M}.\eqno(27)$$

Clearly, as
 stated before, if the collapse is brought to a halt, the states $a^{out\
\dagger}_{+\omega}
 \ket {\Omega_2}$ get
  converted through late ``reflexion" on a bended $r=0$
curve, into a linear
superposition of real quanta on $\cal {I}^+$. The  correlation of these
late non thermal photons
 with earlier Hawking photons is of course maintained and reflects the
purity of the quantum state
 $\ket0$. For genuine collapse, back-reaction is expected to reduce
the total initial mass $M$
to a Planck mass in a retarded time of order $M^3$; of course, the
above computation
 is then at best
 valid for a retarded time $u=O( M^3)$. In both cases Hawking
photons emerge from
 the vacuum through
  production of correlated pairs which should determine the
back- reaction.  We now further analyse the nature of these pairs.
\bigskip \bigskip
\noindent
{\bf 3. Hawking Radiation from Transplanckian Dipoles.}

The Hawking process, in
absence of back reaction is entirely described by the
Heisenberg state $\ket 0$ reexpressed as $ U \ket\Omega$ by Eq.(25).   To
understand the source of
 the back-reaction, it is interesting to uncover from this equation the
detailed history of the
vacuum fluctuations leading to the emission of real quanta. We shall do
this in two steps. First
 we shall use a very simple method to get a qualitative picture of this
history which will then
be made quantitatively precise at the expense of a more sophisticated
formalism.

Let us isolate  the single
pair contributions $\ket p$
 from $\ket 0$  by expanding $U\ket \Omega$ to first order in
$\beta_\omega  / \gamma_\omega$.  Up to a normalization factor we get
$$ \ket p
=\int_\epsilon^\infty d\omega
 \exp (- \omega 4\pi M)  a^{out\  \dagger}_{-\omega}  a^{out\
\dagger}_{+\omega}\ket\Omega \eqno(28)  $$
 where $\epsilon$ is an infrared cut-off. The first
quantized wave-function
$\Psi(u_1,v_1;u_2,v_2) \equiv  \bra \Omega \Phi (u_1,v_1)
\Phi (u_2,v_2)\ket p $ describing $\ket p$ becomes at late times    $$
\lim_{u_2 \to \infty\atop
  v_1 \to \infty }\Psi(u_1,v_1;u_2,v_2) = \Psi(u_1,v_2) =
\int_\epsilon^\infty d\omega
 {\exp (- \omega 4\pi M)\over 4\pi\omega}
 \exp \left[ -i\omega ( u_1+ 4M  \ln {v_2 \over
A}) \right] \eqno(29)$$  where $(u_1,v_1)$
 refers to the Hawking quanta and $(u_2,v_2)$ to its
partner. One can form   2-point conserved
currents out of  $\Psi(u_1,v_1;u_2,v_2)$ and its complex
conjugate but their  time components do not
 define positive definite probabilities; nevertheless one
expects that the wave packets described by
 Eq.(29)   correspond to the localisation of the vacuum
fluctuations at late times and therefore at
 any time via the conservation law. This expectation will
be proven correct in our quantitative method.

The wave-function Eq.(29) of the
correlated
 pair at late times is a sum over frequencies in a range $\Delta
\omega = O(M^{-1})$; it is peaked at $u_1= \bar u$ and $v_2=\bar v$
such that $$\bar u + 4M
\ln { \bar v\over A} =0\eqno(30)$$  and spread over a
range $$\Delta u + 4M
\Delta \ln {   v\over A} \simeq (\Delta \omega)^{-1} = O(M) .
\eqno(31)$$
Comparing the curve Eq.(30)
with the $r=0$ trajectory Eq.(11) we see that the former is the
symmetric of the latter with
 respect to $v=0$ in Fig.2. It is thus a space-like curve which
is readily identified to a
$t=$constant curve, namely $ U(\bar u) + V(\bar v) =2\tau=0$.

We may express the correlated pair
at late times as a superposition
 of pairs of  Hawking photons localized within their wavelength
$$\Delta u_1 = O(M)\eqno(32)$$
correlated to a partner which from Eq.(32) is localized in a
region $$\Delta v_2 = O(\bar v)
 .\eqno(33)$$ Extrapolating back in time, we see that all these
correlated wave packets meet in the flat Minkowski space-time
inside the shell, simultaneously in the Schwartzschild time $t$
along the curve Eq.(30), up to
  spreads Eqs.(32) and (33).   The  local frequency
$\tilde\omega$ of the partner,
in the Lorentz frame fixed by the spherical shell, is obtained from
Eq.(16) by expanding $v$ around
 $\bar v$. Thus $$ \tilde\omega = \omega {4M \over
\vert\bar v\vert}.   \eqno(34)$$
   Similarly, the local frequency of the ``Hawking
fluctuation" itself, that is the
 vacuum fluctuation  generating the Hawking photon  of frequency
$\omega $, is  from Eq.(10) $\omega 4M/   U(\bar u)$, and
thus equals   the partner frequency
 $\tilde\omega$.\foot{Similar conclusions were reached in reference
[12] by examining local Bogoliubov
coefficients.}  Extrapolating further back in time, we can trace the
correlated wave-packets back to $\cal{I}^-$.
 Thus, starting from there, the Hawking fluctuation and
its partner form  concentric spheres traveling
 with the velocity of light and separated at a given
time $t$ by a radial distance $v$ of the order
 of their local wavelength. After crossing $r=0$, the
Hawking fluctuation  merges on the curve Eq.(30)
 with its partner which was chasing it ; the
fluctuations then separate: one member propagates
 towards $\cal {I}^+$ and converts to a real Hawking
photon while its partner propagates towards the  horizon.

The above description
 of the Hawking radiation is expected to apply for retarded times $O(1) <
u_1 <O(M^3)$  and thus
 for $O(M)>\bar v >O(M \exp -M^2)$ which means that the local
frequencies of a pair
goes up from $ \tilde \omega = O(M^{-1})$ to transplanckian values $
\tilde \omega = O(M^{-1}
\exp +M^2)$. Correspondingly the localization is focused for late
retarded time $u_1$ down
 to a radial spread $\Delta r =  O(M\exp -M^2)$.  Planckian
frequencies and distances
 are reached   a very short time after the onset of Hawking radiation,
namely after a time $u_p$ such that $$u_p = O(M\ln M). \eqno (35)$$
More generally one would
get a quantum superposition of such sharply defined  correlated
pairs.

The common local
 frequency of the Hawking fluctuation and of its partner has a deep
significance.
Comparing Eq.(34) to Eq.(10), we see that (the equality is exact for
$\lambda=1$)
$$ { \omega \over \tilde \omega} = \sqrt{g_{00}^{in}(t)} \eqno(36) $$
where $t$ is the
 time at which the Hawking fluctuation emanating   from a point
of the curve Eq.(30)
 crosses the shell. Eq.(36) expresses  then  the redshift
experienced by this
fluctuation when it moves with the velocity of light from inside the
shell to $\cal {I}^+$
 and seem to indicate that the Hawking fluctuation required to make a
Hawking photon must,
on the average, carry the energy necessary to overcome the gravitational
potential energy to
reach infinity. However, the above qualitative considerations do not
explain the relation
 between local frequencies
  and local energies in the vacuum. In particular
the fact that the equality
 between the frequencies of the Hawking fluctuation and of its
partner is consistent with
 the fact that the total Minkowskian energy is zero on $\cal
{I}^-$ remains mysterious.
 To clarify these issues and get a complete picture, we now turn to
a quantitative description
 of the structure and the history of the pair.\foot{The forthcoming
discussion follows the
analysis of  reference [13]
where the energy content of vacuum fluctuations
is expressed in terms of weak
 values$^{[14]}$. Weak values
  were previously used to describe  the creation of a
pair of charged particles in
an external electric field$^{[15]}$ and a clear picture of the
emergence of the pair out of
vacuum fluctuations was obtained in this way.}

Let us consider a normalized
 state describing a Hawking photon on $\cal {I}^+$ localized on the size
of its wavelength; its state vector is
$$ \ket {P_1} = \int_0^\infty
 d\omega f(\omega)  a^{out\  \dagger}_{-\omega} \ket{\Omega_1}
\eqno(37)$$ where $f_(\omega)$
 is a complex function whose modulum is centred around a
frequency $\omega_0$ of order
$M^{-1}$ and spreads over a comparable range. This state is EPR
correlated to $ \bra{P_1}0\rangle$
 and the pair can be represented, up to a normalization
factor by  $\ket{P_1} \bra{P_1}0\rangle$. Using Eq.(25), we have
$$\ket{P_1} \bra{P_1}0\rangle =
 \bra \Omega 0 \rangle  \int d\omega d\omega^\prime
 {\beta_\omega \over \alpha_\omega}   f^*(\omega)
f(\omega^\prime)  a^{out\  \dagger}_{-\omega^\prime}  a^{out\
\dagger}_{+\omega}\ket{\Omega_1}\ket{\Omega_2}. \eqno(38)$$
The matrix element of the energy momentum tensor
operator $\hat T_{\mu \nu}(x)$
between the vacuum state $\ket 0$ and the correlated pair,
suitably normalized, is called
the weak value of the operator for the post selected state $\ket
{P_1}$, namely
$$ T_{\mu\nu}^{weak}(x)\equiv{\langle 0\ket{P_1} \bra{P_1}\hat T_{\mu
\nu}(x)\ket 0 \over
 \langle 0\ket{P_1} \bra{P_1}\ket 0}. \eqno (39)            $$
Aharonov and al have shown that, if a future
measurement were to
 yield the  post
  selected state,   the real part of the weak value of a hermitian
operator     can
 be recorded  by a ``weak" non demolition measurement and  its
imaginary part
induces a shift
 of the conjugate
  variable of the measuring device$^{[14]}$.

By post selecting
 a rare event one
  can gain information about quantum fluctuations which are
averaged out in
 expectation values. In this way Eq.(39)  selects out of the full wave
function the
 contribution
  of the pair
   considered and   provides the quantitative elements lacking in
our previous
 description.
  It yields
   indeed the
    values of the  energy-momentum tensor hidden in the
vacuum fluctuations
 needed to generate
  the final state $\ket {P_1}$ on $\cal {I}^+$ out of the
initial state $\ket 0$.

$\hat T_{\mu \nu}(x)$
 operating on $\ket 0$ is a sum of a multiple of the unit operator and
a bilinear form in $a^{in\ \dagger}$. It follows from  Eq.(20) that
$$  a^{in\  \dagger}_{-\omega^\prime}  a^{in\
\dagger}_{+\omega}\ket 0 ={1\over \alpha_\omega \alpha_{\omega^\prime}}
 a^{out\  \dagger}_{-\omega^\prime}  a^{out\
\dagger}_{+\omega}\ket 0
- {\beta_{\omega }\over \alpha_{\omega }} \delta (\omega -
\omega^\prime) \ket 0 \eqno(40) $$
so that it can also be
 expressed as a sum of
  a multiple $-\beta_\omega / \alpha_\omega$ of the unit
operator and a bilinear
 form in $a^{out\ \dagger}$. From the definition Eq.(39)  of
$T_{\mu\nu}^{weak}(x)$,
 we see that the unit
  operator yields a contribution independent of the post
selected state $\ket {P_1}$.
 We therefore define
  $$  \tilde T_{\mu\nu}^{weak}(x)=
    {\langle 0\ket{P_1} \bra{P_1}\hat T_{\mu
\nu}(x)\ket 0 \over \langle 0\ket{P_1}
 \bra{P_1}\ket 0} - {\bra \Omega \hat   T_{\mu \nu}(x)\ket
0   \over \bra \Omega 0 \rangle    }\eqno(41)  $$
which does not depend on the second term
of Eq.(40) and truly characterizes the pair. Note that
$\tilde T_{\mu\nu}^{weak}(x)$ is  independent
 of the subtraction needed to renormalize  the
vacuum expectation value of $\hat T_{\mu \nu}(x)^{[15]}$.

We first compute on $\cal {I}^+$ $\tilde
 T_{uu}^{weak} $ which measures the energy density
carried by the Hawking photon. From Eq.(38),
 the weak value Eq.(39) is given by
$$ T_{\mu\nu}^{weak}(x) ={\int d\omega d\omega^\prime {\beta_\omega \over
\alpha_\omega} f(\omega) f^*(\omega^\prime) \bra \Omega a^{out}_{+\omega}
a^{out }_ {-\omega^\prime}\hat T_{\mu \nu}(x)\ket 0  \over \int d\omega
\left[{\beta_\omega \over \alpha_\omega}\right]^2
 \vert f(\omega)\vert^2  \bra \Omega
0\rangle}.\eqno(42 )$$
Using
 Eqs.(40)
  and (41) and the energy momentum tensor $\hat T_{\mu \nu}(x)$ of
the rescaled
 field $\Phi$ (corresponding to an energy density    $\hat T_{\mu
\nu}(x) / 4\pi r^2 $), we get
$$ \lim_{v \to +\infty}\tilde T_{uu}^{weak}=
{{1\over 2\pi}\int d\omega d\omega^\prime \left[{\beta_\omega \over
\alpha_\omega}\right]^2 f(\omega)
f^*(\omega^\prime)\sqrt{\omega\omega^\prime}
\exp [-i(\omega-\omega^\prime)u]
\over  \int d\omega  \left[{\beta_\omega \over
\alpha_\omega}\right]^2 \vert
f(\omega)\vert^2   }\eqno(43)  $$ and $$\int du
\lim_{v \to +\infty}\tilde
T_{uu}^{weak} = <\omega>= { \int d\omega  \left[{\beta_\omega \over
\alpha_\omega}\right]^2 \vert f(\omega)\vert^2 \omega\over
 \int d\omega
   \left[{\beta_\omega \over \alpha_\omega}\right]^2
   \vert f(\omega)\vert^2
}.    \eqno(44)$$
Eq.(44) expresses that the total average energy of the
 Hawking photon is $<\omega>$. Note that
the average is taken not only with respect to the quantum
 weight $f(\omega)$ but also over a
thermal distribution at the Hawking temperature $(1/ 8\pi M)$.  The
distribution is here maxwellian because we have  post selected a single
pair; this is an interesting result which deserves further
 analysis$^{[13]}$.  Eq.(43) give the local
(complex) energy content of the quantum and we learn that
to localize it around a retarded
time $u_0$ on its wavelength scale we must choose the phase
 of $f(\omega)$ accordingly, for instance
$$f(\omega)=\vert f(\omega) \vert \exp (i\omega u_0). \eqno(45)$$
We now evaluate
 the incident
  distribution of vacuum fluctuation energies building the Hawking
quantum and its
 partner. To this
  effect we compute  $\tilde T_{vv}^{weak} $ on $\cal {I}^-$,
taking into account Eq.(45). We get
$$ \lim_{u \to -\infty}
\tilde T_{vv}^{weak}={{1\over 2\pi}\int d\omega d\omega^\prime
 {\beta_\omega \over
 \alpha_\omega}{\beta_{\omega^\prime} \over \alpha_{\omega^\prime}} \vert
f(\omega) f^*(\omega^\prime)
\vert {4M \sqrt{ \omega\omega^\prime}\over (v+i\epsilon)^2} \exp
\{i(\omega-\omega^\prime)[u_0
+ \ln {v+i\epsilon\over A}]\}       \over  \int
d\omega  \left[{\beta_\omega
\over \alpha_\omega}\right]^2 \vert f(\omega)\vert^2}\eqno(46) $$
We see
 that
 $\lim_{u \to -\infty}\int dv
 \tilde T_{vv}^{weak}$ is zero. From (41), we verify that
this must indeed be the case
 because $\ket 0$ is an eigenstate of total energy with eigenvalue
zero.  We also see that for $v>0$, $\tilde T_{vv}^{weak}$ is real
and positive and therefore
that the  locally complex contribution for $v<0$, which is the
Hawking fluctuation,  must
integrate to a negative real value. We thus have two overlapping
wave packets located around
 $v_\pm=\pm A\exp(-u_0/4M)$, that is at the values  of $v$
corresponding to the intersection
 of the ray travelling along $u=u_0$ and of the curves $r=0$
and $\tau =0$, Eqs (11) and (30).
 Their spread is of order $\vert v_\pm \vert = \vert \bar v
\vert$ and they carry an energy of
 the order of $ \tilde\omega = 4M \omega_0 / \vert\bar v\vert$ as
in Eq.(34). It is easy to verify, comparing $\tilde T_{vv}^{weak} $ and
$\tilde T_{UU}^{weak} $ that these
 wave packets travel in empty space with the velocity of light,
keeping their shape. The qualitative
 behaviour previously described is entirely recovered but we now
have a precise picture of the energy
 densities carried by the fluctuations. It remains to understand
how the negative total energy carried
 by the Hawking fluctuation on ${\cal I}^-$ gets converted into a
positive energy photon on  ${\cal I}^+$.

We see from Eq.(46) that the post selection
 of a monochromatic photon by a
$\delta$-function peaking $\vert f(\omega)\vert$
 at $\omega_0$, would render   $\tilde
T_{vv}^{weak} $ real
 and positive both for
  positive and negative $v$, leaving   a negative energy
singularity at $v=0$ to
 restore the vanishing of the total energy of the pair.\foot{Such a
singular behaviour for
the expectation value of
 $T_{\mu\nu}$ is encountered in the Fulling-Rindler
vacuum.$^{[16]}$} This
singularity is avoided in the normalized wave packet formulation and the
resulting behaviour is illustrated in Fig.3. The negative (and
the imaginary) contributions
 to the Hawking fluctuation  are mostly  concentrated in the oscillatory
tail of the wave packet near $v=0$,
 or after reflection, at large $u$. The ``core" of the wave
packet, centered at $- \bar v$ with local frequency
 $ \tilde \omega =4M \omega_0 /  \vert
 \bar v \vert$ remains real and positive.
The negative tail energy overcompensates
 the positive energy of the core.  The conversion of the
vacuum Hawking fluctuation depicted in Fig.3b
 to the Hawking photon depicted in Fig.3a is achieved by
the differential redshift $ du/ dU = -4M/ U$.
 This redshift converts the frequency $ \tilde \omega$
of the core to $\omega$ and damps exponentially
 the oscillatory tail near $v=0$ (while enhancing the
oscillations on the other side of the core).

The differential redshift  encodes the loss of core energy in the
gravitational background outside the star. In
 addition it changes the sign of the total energy of
the Hawking fluctuation and may be viewed as the detailed
mechanism underlying pair production through
  frequency sign shift in  the  Bogoliubov transformation.
This can made even more explicit
 by using local Bogoliubov transformation$^{[17]}$ to interpolate
smoothly between the $U$-description
 inside the star to the $u$-description on ${\cal I}^+$.

We now have a complete description for
 the emission of a (s-wave) Hawking quantum. At
$t=-\infty$, on $\cal {I}^-$, we select
 among the vacuum fluctuations of empty Minkowski
space, a dipolar spherically symmetric
distribution with total energy zero, moving with light
velocity, carrying positive energy on the
 outer pole and such that the inner pole carries a
total negative energy but with a positive
 energy core. The outer pole   never reaches $r=0$ for
the external observer. After passing the
curve $r=0$ the inner pole meets the outer one and then
separates,
propagates
towards  $\cal {I}^+$ and converts to a real quantum of positive
energy by
 loosing its negative
  tail and redshifting its positive core.  The real quantum is
still correlated in an
 EPR fashion with the positive energy fluctuation moving towards an
eventual horizon. The
local frequency $\tilde \omega$ of the dipole pair is determined by the
redshift required to
 realize the conversion
  of the core of the Hawking fluctuation into a real
quantum, or equivalently
 by energy conservation. The distance between the two poles before
their separation is of
order $\tilde \omega^{-1}$
 and hence comparable to their spread. For
quanta emitted a very short
 time  after the start of the Hawking emission, namely after a time
$t_p$ given by Eq.(30),  the
 dipole ancestors acquire very large transplanckian frequencies and
are localized over  very short
 cisplanckian distances. This
 remains true if after a time long
compared to $t_p$ the geodesic
 collapse is brought to a halt. The only difference is that now
the outer pole does reach the
bended $r=0$ curve and then  propagates towards $\cal {I}^+$
giving birth to a superposition
 of quanta whose correlations ensure the pure state character
of the final radiation state.

Although transplanckian dipoles
 are totally consistent in the framework of the free field
theory used here, their use in a
 theory where gravitational back-reaction should be included
is puzzling and raises   many questions.  We now discuss these points.
\bigskip
\noindent
 {\bf 4. Gauge invariance, back-reaction and the unitarity issue.}

The gravitational interaction
 experienced by transplanckian dipoles in the empty Minkowski
background space cannot be
 deduced from Einstein equations because  of the
non renormalizability of
the quantized theory: we do not know the effective coupling at such
small distances and at the
 high energy densities encountered in the dipoles histories, nor do
we understand how to depart
 from a non fluctuating classical background metric. Therefore the
   pair production mechanism
   in presence of these  gravitational nonlinearities
 seems to escape our ken, and
  so does therefore the back-reaction they induce on the
metric. The fact that the dipoles
 are part of vacuum fluctuations does not help: it
simply means that although they would
 not contribute to expectation values of the energy, they
will enter general matrix elements, as
 for instance in the evaluation of the weak value $T_{\mu
\nu}^{weak}(x)$ or in the vacuum expectation
 values of correlators $\langle 0 \vert\hat
T_{\mu\nu}(x) \hat T_{\sigma\tau}(y) \ket 0$.

It is important to realize that it is impossible to reduce
transplanckian dipoles to planckian ones by some gauge
transformation  as long as a  background exists.
 One would have    to scale
$\tilde\omega$ down to the Planck scale everywhere
 on the dipole spheres and between them so that
their separation $ \tilde\omega^{-1}$ would be stretched
 accordingly to a Planck lenght. This cannot
be achieved through local Lorentz transformations without
 reintroducing transplanckian Unruh
frequencies. Global Lorentz transformatioms are not available
 either. Indeed, in the extended
coordinates $(u,v)$ which can be  used by the distant observer,
  the frequency $\tilde\omega$
depends only on the Lorentz boost parameter  $\lambda$ in Eq.(5)
 which is fixed from the continuity
equation for $r$ at the star boundary (it is equal to one in the
 limit of a light-like collapse).
Thus  the invariant meaning of $\tilde\omega$ is explicitly related
 to the existence of the collapsing
star. This should be contrasted with    the
dependence of Minkowskian frequencies on   Lorentz boosts in the
 description of the Hawking
radiation from the Unruh vacuum where  the effect of the star collapse
 is mimicked by suitable
boundary conditions on a fictitious past horizon. The latter description
 introduces a spurious
symmetry  which is just such a Lorentz boost   and which is equivalent
 to a global Killing symmetry $t
\to t+$constant. This global symmetry is clearly broken inside the
 collapsing star.

We now examine whether the transplanckian frequencies can be reduced
 dynamically, in absence of
gravitational back-reaction, by using more realistic field theoretical
 models. We shall argue
that this is not possible because the
s-wave transplanckian frequencies are for free fields part of a
 thermal bath containing higher
angular momenta at the same local temperature $ T_{loc}$ and that
 this thermal
bath
 at transplanckian temperatures survive   conventional renormalizable
interacting field theories

The free scalar field radial partial wave equation
$$ {\partial^2 \Phi^{(l)} \over\partial t^2} - {\partial^2 \Phi^{(l)}
 \over\partial
r^{*2}}-\left(1-{2M\over r}\right)\left({2M\over r^3} + {l(l+1)\over r^2}
\right) \Phi^{(l)}=0
\eqno(47) $$
illustrates  that, outside the star, at coordinate distances
$$\eta \equiv (r-2M) << 1\eqno(48) $$ of the
horizon, the centrifugal barrier for the radial partial wave of
 angular momentum $l$,   centred
at $r=3M$,   goes down  as $\eta l(l+1)/8M^3$. This is an exponential
 drop in $r^*$, and thus
for a mode of frequency $\omega$, $\Phi^{(l)}$ will, outside the
 star, propagate
roughly as a free wave for coordinate distances   smaller than
 $\eta (l)$ given by
$$\omega^2 = {\eta l(l+1)\over 8M^3}.\eqno(49)$$
 Outside the star, in the immediate vicinity of the shell, the
  local frequency is $   \omega
\sqrt{2M/\eta}$ and,   from Eq.(9), $\Phi^{(l)}$ propagates
 inside the star at a still higher
frequency  $   \omega 2M/\eta$. It is  therefore quite
insensitive to the centrifugal barrier $l(l+1) / r^2$ there
 until it reaches very small
distances from the centre of the star ( $r \leq \sqrt{2\eta M}$)
 whereupon it is
reflected.  Thus the analysis of sections 2 and 3 is
essentially valid for all angular momenta except that most of the
 high angular momentum modes are
reflected outside the star at a radius $\eta (l)$   towards the horizon.
Transplanckian dipoles exist for all angular momenta.   Outside
 the shell the distribution of
Hawking photons ancestors of frequency $   \omega \sqrt{2M/\eta}$
 is the result of a balance of
outgoing and reflected ``hot photons".  They are   nearly
in thermal equilibrium close to the horizon outside the star$^{[18]}$
  with a local blueshifted
temperature $$ T_{loc} = {1\over 8\pi M}(1-{2M\over r})^{-1/2} =
 O({1\over \sqrt{\eta M}}).\eqno(50)$$

Let us explain qualitatively how this picture arises. The number
 of hot photons $N$ in the
thermal distribution Eq.(50), crossing a sphere of radius $\eta$
  per unit Schwartzschild time
is roughly
$${dN\over dt}  = O\left[ M^2 T_{loc}^3(\eta)
\sqrt{\eta \over M}\right]\eqno(51)$$
where the last factor measures the dilation from
 local time to time at infinity. The
ratio $\rho$ of such modes   reaching $\cal {I}^+$
to the number of reflected modes with $\omega =
O(1/M)$ is of order $l^{-2}$ where $l$ is given by
Eq.(49). Thus   $$\rho = O\left[
{\eta \over M}\right].\eqno(52)$$ Each mode passing
 the barrier carries to infinity an energy of order
$1/ M$, so that the total amount of radiated energy per unit time is
$${dM\over dt}=  O\left[ M T_{loc}^3(\eta)
\sqrt{\eta \over M}\rho\right]=O\left[{1 \over
M^2}\right].\eqno(53)$$
This result describes correctly the radiated
 flux but the important point is that the
dimensional parameter $\eta$ has canceled out in Eq.(53),
expressing the fact that the total
flux at infinity can be estimated from any sphere
sufficiently close to the horizon by
replacing the sphere by heat source of hot photons
 at the local temperature Eq.(50). Note that the
thermal cloud of hot photons is consistent with the
 finite limit of the expectation value of the
energy on the horizon because of the compensating
 divergence of the negative Schwartzschild vacuum
energy.

The inclusion of higher angular momentum for free fields does not change
the transplanckian character of the production
 process. Rather, it imbeds, outside the
star, the s-wave transplanckian frequencies in
   a transplanckian thermal bath, which at this
stage is essentially kinematical as it does not rely on
interactions. Interactions due do
asymptotically free interacting renormalizable field theory,
  mixing different angular momenta
of single hot photons will be weak at the Planck energy and
 could only contribute to stabilize
the temperature.  Hence, as stated above, they cannot
 reduce the frequencies
$\tilde\omega$ to planckian or cisplanckian values.

Thus, to describe correctly the vacuum
fluctuations responsible for the Hawking process,
 gravitational interactions must be taken
into account  at energies and distance scales
  where the classical theory does not seem to make
any sense. In other words, understanding correctly
 the production of Hawking photons requires
at least some genuine properties of quantum gravity.

At this point, one might question
 whether the Hawking radiation should at all exist: a Planck
scale cut-off of vacuum fluctuations
 would clearly wipe out production of Hawking quanta
except for the very few emitted before
the time $t_p$. We believe that this would be an
unreasonable conclusion. In presence of an horizon, the
Hawking radiation appears indeed to have
 thermodynamical significance. If one admits the
Bekenstein conjecture that the area of the
 event horizon is a measure of entropy, then the
black hole entropy must be, for dimensional
 reasons, inversely proportional to the Planck
constant. This in turn requires that an eternal
 black hole should have a global temperature
proportional to $\hbar$. Consider indeed the classical
 Killing identity$^{[19]}$ (it can
be viewed as the integrated constraint equation over a
 static coordinate patch) which can be
written as  $$-{\kappa \over 2\pi} \delta {A\over 4}=
\delta H -\delta M_\infty
\eqno(54)$$
where $\kappa= 1/4M$ is the surface gravity of the hole
 of mass $M$, $M_\infty$ the total
mass at infinity   and $\delta H$ is the variation of all
 non gravitational parameters in the
matter hamiltonian outside the horizon. The Bekenstein
 conjecture   implies that
there exists a global temperature
proportional to the surface gravity. But (54) being a
classical equation, this temperature
should be proportional to $\hbar$ to cancel the
 $\hbar^{-1}$ in the entropy. This
gives credence to the estimate of this
 temperature via euclidean continuation of
the metric, either for Green's functions,
 for partition functions$^{[20]}$ or
for tunneling amplitudes$^{[5]}$, because
 euclidean continuation always leads to
the required dependence on $\hbar$ and because
 all these methods yield the same
result, namely the Hawking temperature  Eq.(27). The fact that the
thermodynamical argument refers more directly
to hypothetical eternal black
holes than to incipient one does not weaken its significance. The pair
production mechanism in an incipient black hole
 results as shown above in a
local thermalization close to the horizon which
 is consistent with the global
Hawking temperature and this is presumably a
 general feature. In other words,
close to its  horizon, incipient black  holes
 tend to behave as eternal ones.
The thermodynamic significance of the Hawking
 result Eq.(27) suggest that its
derivation through the dynamics of free field,
 thus in absence of gravitational
interactions, is a particular realisation of a
 more general phenomenon and  that
the Hawking radiation is  a necessary concomitant
 of a geodesic collapse.

The taming of the transplanckian dipoles and the related transplanckian
local temperatures poses then a fundamental problem whose solution
apparently does not lie in conventional physics.
Despite our ignorance,  a quest on how to
achieve this  without a trivial cut-off at
 the Planck scale may provide useful clues for
entering the unsafe land of quantum gravity.

We shall argue that
such a taming mechanism can be found in a
general feature of the closed string theory approach
to
gravity, independent of the particular model
 used and in fact independent to a
large extend to the detailed structure of the
 string theory itself in the limit of weak coupling.
This is the reason why we shall restrict our
considerations to this limit although interactions may
lead to interesting consequences$^{[21]}$ but
 rely on more detailed aspects of string theories
whose theoretical foundations are at best incomplete.
 The assumption of weak coupling means that a
gas of strings can be approximatively described by a
 set of free massless and  massive fields with an
exponential asymptotic density of states $\rho(m)$ of
 mass m: $$ \rho(m)= A m^{-D} \exp (\beta_0 m).
\eqno(55)$$  In string theory, the inverse Hagedorn
temperature $\beta_0$ depends,
for a given string
tension, on the left and right central charges and $A$
 depends in addition on the dimensionality of
space-time $D$ $^{[22],[24]}$. The string tension must
 be chosen such  that  $\beta_0$  is larger
than unity.

Such weakly interacting closed strings in four dimensional
 flat space-time exhibit, above a critical
energy density, a phase transition whereby any additional
increase of density in the massless modes
would condense at fixed  temperature into infinite strings
 of classical Hausdorff dimension
two$^{[22],[23]}$.  As explained below, these results rely
 on the value of the prefactor
$m^{-D}$, as always in the case of exponential energy
spectrum$^{[25]}$.  It is therefore
important to realize that the exponent $-D$ is independent
of the particular closed string theory
used (provided the theory does not allow open strings) and
even of the particular compactification
scheme. In fact it requires only one property of closed string
 theory: namely that the classical
lenght of massive free closed strings has the shape of a random
 walk; one can indeed verify that the
form of Eq.(55), including the value $(-D)$ of the exponent in
the prefactor, follows from the
central limit theorem as applied to closed random walk$^{[26]}$.

We briefly review how the phase transition arises$^{[22]}$.
 Consider the following toy model: a box
of arbitrary large volume $V$ in $D-1$ spatial dimensions
 contains a total energy $E$ shared between
two constituents in thermal equilibrium; a gas of massless
 particles and a macroscopic ``string" of
finite energy density characterized only by a density of
states growing with its energy $E_s$ as
$\exp (\beta_0 E_s)$. Denoting by $S_g$, $S_s$, $S_{g+s}$
 respectively the entropies of the gas, the
string, and the string-gas system, we have
 $$\eqalignno{ &S_g = {D\over D-1} V^{\prime\, 1/D}
E^{(D-1)/D} &(56)\cr &S_s= \beta_0 E &(57) \cr
&S_{g+s}= {D\over D-1} V^{\prime\, 1/D}
 (E-E_s)^{(D-1)/D} + \beta_0 E_s. &(58)}$$
Here $V^\prime = \xi V$ where $\xi$ is a
 number and the temperature of the gas is $\beta^{-1} =
[(E-E_s)/V^\prime]^{1/D}$. Equilibrium of the two phase system
 implies $\beta = \beta_0$ and it is
a stable one. Thus one can rewrite Eq.(58) for $E > \beta_0^{-D}
V^\prime$ as
$$S_{g+s}=\beta_0 E + {1\over D-1} \beta_0^{1-D} V^\prime .\eqno(59)$$
Following Eqs.(56), (57) and (59), we have plotted in Fig.4, as a
 function of the total
energy density $\sigma = E/V^\prime$, the corresponding entropy
 densities $s_g$, $s_s$, and
$s_{g+s}$. Clearly, for $\sigma$ exceeding  $\sigma_c \equiv
 \beta_0^{-D}$, $s_{g+s}$ becomes
greater than $s_g$ and any increase of density in the gas
 above $\sigma_c$ would condense into the
(infinite) string at the Hagedorn temperature $T=\beta_0^{-1}$.

Consider now, in the weak coupling limit,  genuine  strings
  with   total energy $E$ enclosed in the
volume $V$. The main feature ignored in the toy model is the
 existence of a  spectrum of massive
modes extrapolating between the zero mass states and macroscopic
 strings encoded in Eq.(55). If $D$
is large enough ($D \geq 4$), small massive closed strings are
 strongly favoured with respect to
larger ones and the total contribution of finite energy strings
 (in an infinite volume) gives only a
finite contribution to the energy and to the entropy densities
 when $T$ reaches $\beta_0^{-1}$. The
qualitative results of the toy model remains then valid: if
 additional energy is poured into the
system it will condense into  infinite strings at the temperature
 $\beta_0^{-1}$.   Large and
infinite strings are, classically, random walked shaped. Infinite strings
play the role of an entropy reservoir in thermal equilibrium  which
 massless modes and with a
universal distribution of closed strings.\foot{The appearance of
 an infinite string occurs also for
$D=3$ but below $\beta_0^{-1}$ because of large energy fluctuations.}

We have seen previously that vacuum fluctuations of massless fields
 outside the star are well
described by   thermal distributions with local transplanckian
 temperatures when it approaches the
horizon. Let us now assume that the spectrum   contains a large,
 presumably
exponential, degeneracy of massive states. The toy model and the
 theory of weakly coupled closed
strings suggest that entropy considerations would favour a kind of
 condensation of these hot photons
into an extended object similar to a macroscopic string envelopping
 the horizon. Residual photons
would be distributed at the cisplanckian temperature $\beta_0^{-1}$.
 In this way  a hot  thermal
distribution  at a fixed temperature at a fixed distance of the
 horizon, from which the Hawking
photon can be generated as in   conventional  field theory, could
still exist. But the feeding of
this thermal distribution by a still higher temperature at still
 smaller radius could be avoided and
transplanckian frequencies might disappear from the spectrum of
 vacuum fluctuations.

Such a reconditioning of the vacuum close to the horizon  could
   cut off transplanckian frequencies while precipitating extended
    structures whose energy density
is hopefully not transplanckian.  Further
analysis is of course required to inquire into the consistency and
 the stability of such a
condensation in a finite volume with a non trivial metric, but
 progress along these lines
is possible. Note that this scenario could be consistent with weak
 coupling at the Planck
scale as is the case for weakly coupled closed string theories
 which do include gravitons.

This scheme is related to recent conjectures of Susskind$^{[27]}$
 and may be viewed as an attempt
towards formulating in dynamical terms the brick wall model$^{[28]}$
 or the streched horizon
model$^{[7]}$, or alternatively an achronon$^{[5]}$. At this stage,
 it is however far from obvious
that, as   proposed in references [6] and [7], the classical
collapse would be observer dependent. Hopefully, in view of the fact
 that our dynamical approach  may
be consistent with a weak coupling to gravity, one could investigate
 whether or not this strong form of
complementarity$^{[7],[27]}$ is needed or   if and how a more
 conventional approach to the unitarity
issue, with or without remnants, is still available. \bigskip\bigskip

\centerline{\bf Acknowledgements}
\medskip

We are  grateful to Y.Aharonov, R.Argurio, R.Brout, A.Casher,
  S.Nussinov and S.Popescu for their help
in clarifying the issues addressed to in this paper. One of us
 (F.E.) thanks the Centre for
Microphysics and Cosmology at the Racah Institute  of the Hebrew
University of Jerusalem for their
warm hospitality during the completion of this work. \vfill \eject

\centerline{REFERENCES}

\item{[1]}
S.W. Hawking, Commun. Math. Phys. {\bf 43} (1975) 199.

\item{[2]}
J.D. Beckenstein, Phys. Rev. {\bf D7} (1973)  2333.

\item{[3]}
S.W. Hawking, Phys. Rev. {\bf D14} (1976) 2460 .

\item{[4]}
Y. Aharonov, A. Casher and S. Nussinov, Phys. Lett. {\bf B191} (1987) 51.

\item{[5]}
A. Casher and F. Englert,
 ``{\it Entropy Generation
 in Quantum Gravity and Black Hole Remnants}", in
``String Theory,
 Quantum Gravity
  and the Unification of the Fundamental Interactions" Ed. by M.
Bianchi, F. Fucito,
 E. Marinari, A. Sagnotti, World Scientific (1993).\hfill \break
A. Casher and F. Englert,
 Class. Quantum Grav. {\bf 10} (1993) 2479.\hfill \break
F. Englert and B. Reznik,
 ``{\it Entropy Generation by Tunneling in 2+1-Gravity}" Preprint
TAUP-2102-93, gr-qc/9401010, (1994).

\item{[6]}
G.'t Hooft, in the Proceedings
 of the ``International Conference on Fundamental Aspects of Quantum
Theory" (1992), in Honor of Y. Aharonov's 60th birthday.\hfill \break
C.R. Stephens, G.'t Hooft and B.F.
Whiting, ``{\it Black Hole Evaporation
 without Information Loss"}, Preprint
THU-93/20; UF-RAP-93-11, (1993).

\item{[7]}
L. Susskind, L. Thorlacius and J. Uglum,
 Phys. Rev. {\bf D48}(1993) 3743.

\item{[8]}
J.M. Bardeen, Phys. Rev. Letters {\bf 46} (1981) 382.\hfill \break
R. Parentani and T. Piran, Hebrew University Preprint (1994).

\item{[9]}
T. Jacobson, Phys. Rev.
 {\bf D44} (1991) 1731, {\bf D48} (1993) 728. \hfill \break
K. Schoutens, H. Verlinde,
 E. Verlinde, ``{\it Black Hole Evaporation and Quantum Gravity}"
Preprint CERN-TH.7142/94, PUPT-1441, (1994).

\item{[10]}
W.G. Unruh, Phys. Rev. {\bf D14} (1976) 287.

\item{[11]}
R. Wald, Commun. Math. Phys. {\bf 45} (1975) 9.

\item{[12]}
R. Parentani
 and R. Brout, Nucl. Phys. {\bf B338} (1992) 474.\hfill \break
R. Parentani and R. Brout, Int. J. Mod. Phys. {\bf D1} (1992) 169.

\item{[13]}
S. Massar and R. Parentani, ``{\it Quantum Source of the Back
Reaction:
 Accelerated Detectors and Black Holes}" Preprint ULB-TH 94/02 (1994).

\item{[14]}
Y. Aharonov, D. Albert, A. Casher and L. Vaidman, Phys. Lett.
 {\bf A124} (1987) 199.\hfill \break
Y. Aharonov and L. Vaidman, Phys. Rev. {\bf A41} (1990) 11.\hfill \break
Y. Aharonov, J. Anandan, S. Popescu and L. Vaidman,
     Phys. Rev. Lett. {\bf 64} (1990) 2965.

\item{[15]}
R. Brout,
 S. Massar,
  S. Popescu,
   R. Parentani and Ph. Spindel, ``{\it Quantum Source of the Back
Reaction on a
 Classical Field}" Preprint ULB-TH 93/16, UMH-MG 93/03, (1993).

\item{[16]}
R. Parentani,
  Class. Quantum Grav. {\bf 10} (1993) 1409.

\item{[17]}
S. Massar, R.
 Parentani and
  R. Brout,  Class. Quantum Grav. {\bf 10} (1993) 2431.

\item{[18]}
P. Candelas, Phys. Rev. {\bf D2} (1980) 1541.

\item{[19]}
J.M. Bardeen,
B. Carter and S.W. Hawking, Comm. Math. Phys. {\bf 31} (1973)
161.

\item{[20]}
G. Gibbons and S. Hawking, Phys. Rev. {\bf D15} (1977) 2738, 2752.

\item{[21]}
J.J. Attick and E. Witten, Nucl. Phys. {\bf B310} (1988) 291.

\item{[22]}
Y. Aharonov, F.
 Englert and J. Orloff, Phys. Letters {\bf B199} (1987) 366.

\item{[23]}
D. Mitchell and N. Turok, Phys. Rev. Lett. {\bf 58} (1987) 1577.

\item{[24]}
I. Antoniadis,
J. Ellis and D.V.
 Nanopoulos, Phys. Letters {\bf B199} (1987) 402.

\item{[25]}
S. Frautchi, Phys. Rev. {\bf D3} (1971) 2821. \hfill\break
R.D. Carlitz, Phys. Rev. {\bf D5} (1972) 3231.

\item{[26]}
F. Englert and J.
Orloff, Nucl. Phys. {\bf B334} (1990) 472.

\item{[27]}
L. Susskind, ``Strings, Black Holes and Lorentz Contraction"   Preprint
SU-ITP-93-21, hep-th/9308139 (1993).\hfill\break
L. Susskind,
``Some speculations about Black Hole Entropy in String Theory"
Preprint RU-93-44, hep-th/9309145 (1993).

\item{[28]}
G. 't Hooft, Nucl. Phys. {\bf B256} (1985) 727.

\vfil \eject

\noindent
{\bf Figure Captions}

\noindent
Figure 1. Penrose Diagram of a Collapsing Shell.

The shaded
 region is
  the space-time
   available to the external observer. The dashed lines represent
the  motion of the
 centres of  two
  correlated vacuum fluctuation wave packets; these are the
ancestor of a Hawking
 photon localized on the scale of its wavelength and its partner.

\bigskip\noindent
Figure 2. Collapsing Shell in the $(u,v)$ coordinate system.

The figure depicts the
 shaded region of Fig.1.
  The dashed line reflected on the curve $r=0$ is the
Hawking photon ancestor
and the straight dashed line is its partner. They meet on the curve
$\tau=0$.

\bigskip\noindent
Figure 3. The Vacuum Fluctuation Generating a Hawking Photon.

Fig. 3a represents the
 real part of $\tilde T_{uu}$ on ${\cal I}^+$ that corresponds to a
post-selected Hawking
photon emitted in a
gaussian wave packet centered on $u=u_0$ with frequency
$\omega \simeq (2M)^{-1}$.
 The two crosses on the vertical axis correspond to $\tilde T_{uu} =
\pm {1 \over 4 \pi r^2 (4 M)^2}$.
 The positive energy core is centered around $u=u_0$. Fig. 3b
represents the real part of $\tilde T_{vv}$
corresponding to the same Hawking photon.
The Hawking photon ancestor is located at
negative $v$ (it carries negative total energy) whereas its
partner is located at positive $v$
 (it carries postive energy). Their are oscillations of  $\tilde
T_{vv}$  near $v=0$ for $v<0$ which
 have not been represented. The region
  of positive $\tilde T_{vv}$  on the
left of the drawing centered on $v \simeq -4M e^{- u_0/4M}$ becomes the
positive energy core of Fig. 3a after reflection at $r=0$. The
remaining negative and oscillatory parts of the ancestor correspond to the
oscil
3a. The  crosses on the vertical axis correspond
to $\tilde T_{vv}= \pm {1 \over 4\pi r^2
 (4M)^2}e^{u_0/2M}$. The energy density becomes transplanckian
after a time $u_0=t_p= O(M \log M)$ and is
 localized on a cisplanckian distance $v= \pm 4M
e^{-u_0/4M}$.

\bigskip\noindent
Figure 4. Phase diagram of closed string theory.

The equilibrium
 curve is the dark
  solid line. It coincides with $s_g$ for $ \sigma < \beta_0^{-4}$
and with $s_{g+s}$ for $ \sigma > \beta_0^{-4}$.
\vfill \eject
\end